\documentclass[10pt,twocolumn,showpacs,nofootinbib]{revtex4-1}%
\usepackage[colorlinks,citecolor=blue,linkcolor=blue,urlcolor=blue]{hyperref}
\usepackage{amsmath}
\usepackage{graphicx}
\usepackage{slashed}
\usepackage{xcolor}
\usepackage{parskip}
\newcommand{\punto}{\!\cdot\!}

\begin{document}
\noindent	
\title{QCD determination of the magnetic field dependence of QCD and hadronic parameters}
\author{C. A. Dominguez} 
\affiliation{Centre for Theoretical \& Mathematical Physics, and Department of Physics, University of Cape Town,
Rondebosch 7700, South Africa}
\author{M. Loewe}
\affiliation{Instituto de Fisica, Pontificia Universidad Catolica de Chile,
Casilla 306, Santiago, Chile;}
\affiliation{Centro Científico Tecnol\'ogico de Valparaíso-CCTVAL,
Universidad T\'ecnica Federico Santa Mar\'ia, Casilla 110-V, Valpara\'iso, Chile}
\affiliation{Centre for Theoretical \& Mathematical Physics, and Department of Physics, University of Cape Town,
Rondebosch 7700, South Africa;}
\author{Cristian Villavicencio}
\affiliation{Departamento de Ciencias Basicas, Facultad de Ciencias, Universidad del B\'io-B\'io,
Casilla 447, Chill\'an, Chile}

\date{\today}

\begin{abstract}
\noindent
A set of four finite energy QCD sum rules are used to determine the magnetic field dependence of the sum of the  up- and down-quark masses of QCD, $(m_u + m_d)$, the pion decay constant $f_\pi$,  the pion mass $m_\pi$, the gluon condensate, $\langle \alpha_s\, G^2 \rangle$,  and the squared energy threshold for the onset of perturbative QCD, $s_0$, related to the Polyakov loop of lattice QCD.  
As a first input we take the magnetic evolution of the chiral quark condensate from lattice QCD and/or Nambu--Jona-Lasinio results. 
As a second input we take three different possible conditions concerning the quark and pion masses.
\end{abstract}
\maketitle
\section{Introduction}
\noindent
The method of QCD sum rules (QCDSR) \cite{QCDSR1} is a well-established technique to obtain results in QCD analytically, thus complementing Lattice QCD simulations (LQCD). The extension of QCDSR to finite temperature, as first proposed in \cite{BS}, has contributed significantly to the understanding of hadronic as well as QCD dynamics in this regime \cite{QCDSRT_review}. A further extension of QCDSR to account for the presence of strong magnetic fields was proposed recently in \cite{eB1}.

Modern applications of QCDSR are based on a pioneer proposal  relating QCD to hadronic physics in the complex squared-energy $s$-plane \cite{Shankar}. The only singularities of current correlators  lie on the right-hand plane. They are in the form of poles on the real s-axis (stable hadrons), or on the second Riemann sheet (hadronic resonances). The threshold for the onset of  perturbative QCD (PQCD)  in this plane is named $s_0$, with $s_0 \gtrsim 1\,{\mbox{GeV}^2}$.
\begin{figure}[ht]
\includegraphics[scale=.8]{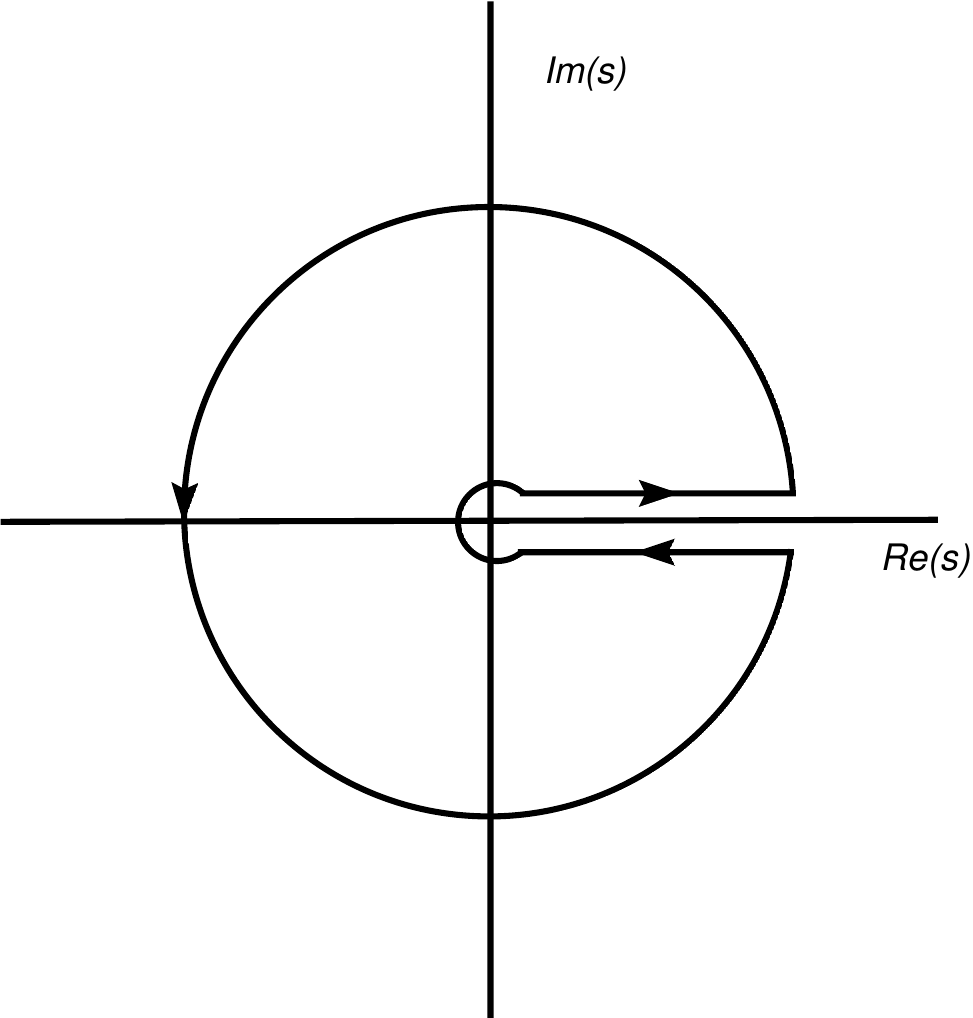}
\caption{\small Integration contour in the complex s-plane. The discontinuity across the real axis brings in the hadronic spectral function, while integration around the circle involves the QCD correlator. The radius of the circle is $s_0$, the onset of QCD.}   
\label{fig:figure1}
\end{figure}
Next, Cauchy theorem is invoked in the s-plane (see Fig.~\ref{fig:figure1}), leading to finite energy QCD sum rules (FESR)
\begin{equation}
\frac{1}{\pi} \int_{0}^{s_0} \!\!\!{\mbox{Im}}\, \Pi(s)|_\text{\tiny{Had}}  \,P(s)\,  ds = 
  \frac{-1}{2 \pi i} \oint_\text{\tiny{C($s_0$)}} \!\!\!\Pi(s)|_\text{\tiny{QCD}}\, P(s)\, ds,  \label{FESR}
\end{equation}
where $P(s)$ is an analytic integration kernel. If singular kernels are needed in applications, Eq.\eqref{FESR} will have to include the residues at the poles.

In the QCD-FESR  framework  at finite temperature \cite{QCDSRT_review}, quark-gluon deconfinement emerges mostly, but not exclusively, from the behavior of $s_0(T)$, as well as other hadronic parameters. For instance, in the light-quark and heavy-light quark systems $s_0(T)$ turns out to be a monotonically decreasing function of temperature, vanishing at a critical value, $T_c$, the deconfinement temperature. It should be mentioned that recently an intriguing connection has been found between $s_0(T)$ and the Polyakov loop, the deconfining object of LQCD \cite{Carlo}. For heavy-heavy quark systems it was first found using QCD sum rules that these states survive the critical temperature for deconfinement \cite{s0HH}. This unexpected situation was later confirmed by LQCD determinations \cite{LQCDHH}.

Another important recent result in this framework is the temperature dependence of the light-quark masses, $(m_u + m_d)(T)$, and the pion decay constant, $f_\pi(T)$ \cite{qmassT}. The latter decreases with increasing temperature, independently of the behavior of the pion mass, as expected from the standard chiral-symmetry scenario in QCD. In contrast, the light quark masses increase substantially with temperature, approaching their {\it constituent quark} values, thus hinting at deconfinement.

Turning to matter in the presence of magnetic fields, a QCD FESR analysis at zero temperature was performed recently in the chiral limit \cite{eB1}. The leading magnetic corrections, quadratic in the field, indicate that $s_0$ is proportional to the absolute value of the quark condensate, $|\langle \bar{q} q \rangle|$. Thus, $s_0$ increases with increasing field, i.e. parameters describing chiral-symmetry restoration behave similarly to those associated with deconfinement. In addition, it was found that the gluon condensate also increases with increasing field strength.

In this paper we improve on the analysis of \cite{eB1} by
considering three current correlators, the light-quark axial-vector current correlator, $\Pi_{\mu \nu}^{AA}$, the hybrid correlator involving a light-quark axial-vector current and its divergence, $\Pi_{5\nu}$, and the correlator of the   divergence of the light-quark axial-vector current, $\psi_5$, in the presence of a magnetic field. The magnetic field will enter as a correction to the propagator in an expansion in powers of $eB$. Two FESR are considered for   $\Pi^{AA}_{\mu\nu}$, with integration kernels  $P(s)=1$ and $P(s)=s$. One FESR is invoked for 
 $\Pi_{5\nu}(s)$ with  $P(s)=1$, and one FESR for $\psi_5(s)$ with $P(s)=1$. This procedure allows for the prediction of four relevant parameters. 
With this choice of correlators and  FESR   there is no contribution from the spin projected quark condensate $\langle \bar q\sigma_{12}q\rangle$.
\footnote{This condensate is otherwise  not negligible in comparison with the standard ones in the vacuum.}    

The sum rules  provide the behavior of the threshold for PQCD, $s_0$, the pion mass, $m_\pi$, the pion decay constant, $f_\pi$, the quark masses, $m_{q}$, and the gluon condensate, $\langle \alpha_s \, G^2\rangle $. One input is required.
One possibility  is to input the  magnetic evolution of the quark condensate, $\langle \bar{q} q \rangle$,  from the Nambu-Jona-Lasinio (NJL) model \cite{Coppola:2018vkw}, which agrees with LQCD results \cite{Bali:2012zg}. 
An  alternative input involves three different possibilities,  (i)
the behavior of the pion mass  is  given by results from  NJL \cite{Coppola:2018vkw}, (ii)
the ratio  $m_q/m_\pi^2$ is assumed constant, with both masses evolving with the magnetic field,  and (iii) the quark masses are assumed to be independent of the magnetic field.\\
Results from this analysis show that $s_0$, and $f_\pi$ always increase with increasing magnetic field, i.e. they are robust quantities. However, the evolution of the gluon condensate is strongly dependent on the assumptions being made for the behavior of the pion and the quark masses.

\section{Vacuum Current Correlators}
\label{sec:vacCorr}
The axial-vector current correlator is defined as

\begin{eqnarray}
\Pi_{\mu\nu}^{AA} (q^{2}) &=& i  \int  d^{4} x \, e^{i q x} \,
\langle 0|T [A_{\mu}(x)  A_{\nu}^{\dagger}(0)]|0 \rangle \, \nonumber \\ 
&=&  q_\mu q_\nu \, \Pi_0(q^2) +g_{\mu\nu}\, \Pi_1(q^2) \; , \label{PimunuAA}
\end{eqnarray}

\vspace{0.5cm}

where $A_\mu(x) =  {:\!\bar{d}(x)\gamma_\mu\gamma_5u(x)\!:}$ is the (electrically charged) axial-vector current,  and $q_\mu$  is the four-momentum carried by the current. 
The functions $\Pi_{0,1}(q^2)$  are free of kinematical singularities, a key property needed in writing dispersion relations and sum rules. Their normalization from the leading order in PQCD is
\begin{equation}
\Pi_0(q^2)|_\text{\tiny PQCD} = - \frac{1}{4 \pi^2} \ln(- q^2/\mu^2) \,, \label{Pi0AA}
\end{equation}
\begin{equation}
\Pi_1(q^2)|_\text{\tiny PQCD} = \frac{1}{4 \pi^2}\, q^2 \ln(- q^2/\mu^2) \,.\label{Pi1AA}
\end{equation}

The operator product expansion (OPE) of current correlators in QCD is given by
\begin{equation}
    \Pi(q^2)|_\text{\tiny QCD} = C_0 \, \hat{I} + \sum_{N=1} \frac{C_{2N} (q^2,\mu^2)}{(-q^2)^{N}} \langle \hat{\mathcal{O}}_{2N} (\mu^2) \rangle \;, \label{OPE}
\end{equation}
where $\langle \hat{\mathcal{O}}_{2N} (\mu^2) \rangle \equiv \langle0| \hat{\mathcal{O}}_{2N} (\mu^2)|0 \rangle$, $\mu^2$ is a renormalization scale, the Wilson coefficients $C_N$ depend on the Lorentz indexes and quantum numbers of the currents, and on the local gauge invariant operators ${\hat{\mathcal{O}}}_N$ built from the quark and gluon fields of the QCD Lagrangian. These operators are ordered by increasing dimensionality and the Wilson coefficients are calculable in PQCD. The unit operator above has dimension $d\equiv 2 N =0$ and $C_0 \hat{I}$ stands for the purely perturbative contribution. The dimension $d\equiv 2 N = 2$ term in the OPE cannot be constructed from gauge invariant operators built from the quark and gluon fields of QCD (apart from quark mass corrections). In addition, there is no evidence for a $d=2$ genuine term  from analyses  using  experimental data \cite{D11,D12}. Hence,   the OPE starts at dimension $d \equiv 2 N = 4$. Quark mass corrections are nonleading in the case of the axial-vector correlator, Eq.\,\eqref{PimunuAA}, and will be neglected in the sequel. 
The contributions at dimension $d=4$ arise from the vacuum expectation values of the gluon field squared (gluon condensate), and of the quark-antiquark  fields (the quark condensate) times the quark mass.

While the Wilson coefficients in the OPE, Eq.(\ref{OPE}) can be computed in PQCD, the values of the vacuum condensates cannot be obtained analytically from first principles, as this would be tantamount to solving QCD analytically and exactly. These condensates can be determined from the QCDSR themselves, in terms of some input experimental information, e.g. spectral function data from $e^+ e^-$ annihilation into hadrons, or hadronic decays of the $\tau$-lepton. Alternatively, they may obtained by LQCD simulations.
An exception is the value of the quark condensate which is related to the pion decay constant through the Gell-Mann-Oakes-Renner (GMOR) relation \cite{GMOR1,GMOR2}, a QCD low energy theorem.\\

The nonperturbative power corrections  for $\Pi_0(q^2)$ are given in terms of the  gluon  and the quark condensates
\begin{equation}
        \Pi_0(q^2)|_\text{\tiny NPQCD} = 
            \frac{1}{q^4} \left[  m_{ud} \, \langle \bar{q} q \rangle
            +\frac{1}{12\pi} \,\langle \alpha_s \,G^2 \rangle  \right], 
            \label{Pi0AANP}
        \end{equation}
where $G^2 \equiv G^{\mu\nu}  G_{\mu\nu}$,  $\langle \bar{q} q \rangle \equiv \langle\bar{u} u \rangle  = \langle \bar{d} d\rangle$, vacuum isospin symmetry breaking will be neglected in the sequel, and
\begin{equation}
m_{ud} \equiv (m_u + m_d) \,.\label{mud}
\end{equation}
Recent values of these quantities are $\langle \alpha_s 
\,G^2 \rangle = 0.037\pm 0.015 \, {\mbox{GeV}^4}$ \cite{GG}, $m_{ud}(2\, \mbox{GeV}) = 8.2 \, \pm \, 0.4 \,{\mbox{MeV}} $ \cite{mud}, and $\langle \bar{q} q \rangle(2\, \mbox{GeV}) = - \, (267 \pm 5 \, {\mbox{MeV}})^3$ \,\cite{GMOR2}. \\

The second current correlator to be considered is $\Pi_{5 \nu} (q^2)$,  involving an axial-vector current and its divergence
\begin{eqnarray}
\Pi_{5 \nu} (q^{2}) &=& i \int d^{4} x \, e^{i q x} \,
\langle 0|T\,[ i\partial^\mu A_{\mu}(x) \,  A_{\nu}^{\dagger}(0)]|0 \rangle  \nonumber \\ [.3cm]
&=& q_\nu\, \Pi_5(q^2)  \; .\label{Pi5DAA}
\end{eqnarray}

\vspace{0.5cm}

In contrast to $\Pi_{\mu\nu}^{AA} (q^{2})$, where quark-mass terms are nonleading in PQCD, in this case they are explicit through $\partial^\mu A_{\mu}(x) =m_{ud} \,{:\!\bar{d}(x)i\gamma_5 u(x)\!:}$. 
The QCD expression for $\Pi_5(q^2)$ up to order $1/q^{2}$ is given by

\begin{equation}
    \Pi_5(q^2)|_\text{\tiny QCD}= - \frac{3}{8 \pi^2} \, {m_{ud}}^2\, \ln(-q^2/\mu^2) +
    2\frac{m_{ud}}{q^2}  \langle \bar{q} q \rangle . 
    \label{Pi5QCD}
\end{equation}

The third current correlator is 
\begin{equation}\label{Psi5}
    \psi_5(q^2) = i\int d^4x\,e^{iqx}\langle 0|T\,[\partial^\mu A_\mu(x)\,\partial^\nu A^\dagger_\nu(0) ]|0\rangle\,.
\end{equation}

Its QCD expression  to order $1/q^{2}$ is given by
\begin{eqnarray}
    \psi_5(q^2)|_\text{\tiny QCD} &=& - \frac{3}{8 \pi^2} \, {m_{ud}}^2\,q^2 \ln(-q^2/\mu^2) +
 \nonumber \\ [.3cm]
 && -\frac{1}{8\pi} \, \frac{{m_{ud}}^2}{q^2}\, \langle\alpha_s\, G^2 \rangle\,
 +\frac{1}{2}\frac{m_{ud}^3}{q^2}\langle \bar{q} q \rangle . \label{Psi5QCD}
\end{eqnarray}

If the axial-vector current correlator, Eq.(\ref{PimunuAA}), were to be written instead  in terms of transverse and longitudinal components,  the longitudinal part would be related to $\Pi_5(q^2)$ through a Ward identity. Also,  $\Pi_5(q^2)$ is related to $\psi_5(q^2)$ also through a Ward identity.
Hence, the use of one FESR for $\Pi_5(q^2)$ and one for $\psi_5(q^2)$, with kernel $P(s) = 1$, is equivalent to the use of two FESR for $\Pi_5(q^2)$, with integration kernels $P(s) = 1$ and $P(s) = s$. However, 
in the presence of a magnetic field the Ward identities are modified as shown below.  Thus,   both correlators will be used,  instead of a single one involving two FESR.\\

To complete the information on the current correlators,  their hadronic representation involves the  lowest state, i.e. the pion

\begin{equation}
\mbox{Im} \,\Pi_0(q^2)|_\text{\tiny HAD} = 2 \pi\,  f_\pi^2\, \delta (q^2-m_\pi^2) 
\label{ImPi0}
\end{equation}
where  $f_\pi = 92.28 \pm 0.07 \, {\mbox{MeV}}$ and the charged pion mass\footnote{
 Hereafter we will refer $f_\pi$ and $m_\pi$ to the charged pion decay constant and charged pion mass, respectively}
 $m_\pi = 139.57018 \pm 0.00035\text{ MeV}$ \cite{PDG}. The next hadronic state, the $a_1(1260)$, with full width  $\Gamma_{a_1} = 250 - 600 \, {\mbox{MeV}}$ \cite{PDG} can be safely neglected, as it lies  above the threshold for PQCD, $s_0 \simeq 1 \, {\mbox{GeV}^2}$, and  its width is quite large  in comparison with the zero-width of the pion. This situation would still prevail  even if $s_0$  grows somewhat in the presence of a magnetic field.\\

The hadronic spectral function for the other two correlators, $\Pi_5(q^2)$ and  $\psi_5(q^2)$ is given by
\begin{equation}
\mbox{Im} \,\Pi_5(q^2)|_\text{\tiny HAD} = 2 \pi \,f_\pi^2 \,m_\pi^2\, \delta (q^2-m_\pi^2)\,,\label{ImPi5}
\end{equation}
and
\begin{equation}
\mbox{Im} \,\psi_5(q^2)|_\text{\tiny HAD} = 2 \pi\, f_\pi^2 \,m_\pi^4 \,\delta (q^2-m_\pi^2)\,.\label{ImPsi5}
\end{equation}

\section{QCD Finite Energy Sum Rules in Vacuum}
\label{sec:vacQCD}

We consider the FESR,  Eq.(\ref{FESR}),  involving $\Pi_0$, Eqs.(\ref{Pi0AA}), (\ref{Pi0AANP}),  (\ref{ImPi0}),  and  $\Pi_5$, Eqs.(\ref{Pi5QCD}),  (\ref{ImPi5}),  and $\psi_5$,  Eqs. (\ref{Psi5QCD}),  (\ref{ImPsi5}). The resulting four  FESR (in vacuum)  are

\begin{align}
2 \, f_\pi^2 
    &= \frac{s_0}{4 \, \pi^2}\,, \label{FESR01}\\
2 \, f_\pi^2 \, m_\pi^2 
    &= \frac{s_0^2}{8\, \pi^2} \,
        -\, m_{ud} \,\langle \bar{q} q \rangle \,
        - \, \frac{1}{12\pi}\, \langle \alpha_s \,G^2 \rangle \,,  
        \label{FESR02}\\
\frac{2 \, f_\pi^2 \, m_\pi^2 }{m_{ud}}
    &= - \, 2\langle \bar{q} q \rangle 
        +\frac{3}{8\pi^2}m_{ud}s_0\, ,\label{FESR03}\\
\frac{2f_\pi^2 \, m_\pi^4}{{m_{ud}}^2} \,
    &= \, \frac{3 \, s_0^2}{16 \, \pi^2} \,-\, \frac{1}{2}m_{ud} \, \langle \bar{q} q \rangle \, +\, \frac{1}{8\pi}\,\langle \alpha_s \,G^2 \rangle \,,
\label{FESR04}
\end{align}
where higher order quark-mass corrections of order $( {m_{ud}}^2/s_0)$ were neglected.
Notice that Eq.~(\ref{FESR03}) is the GMOR relation \cite{GMOR1,GMOR2}, including a higher order quark-mass correction, i.e. $\cal{O}$$(m_q^2)$.\\

We use as an input the charged pion mass and pion decay constant, in order to obtain, as a result, all the other parameters.
These results are needed only as the vacuum normalization of the  magnetic field behavior of the various QCD and hadronic parameters. In other words, they are not to be considered as predictions of their  vacuum values. The reason being that no hadronic states beyond the pion are included. These states are the $a_1(1260)$ and the broad pionic resonances $\pi(1300)$, $\pi(1800)$, lying well beyond the integration range of the magnetic FESR .

\bigskip
\section{Current Correlators in an external magnetic field}

The presence of an external magnetic field  modifies  current correlators in several ways.
First is the minimal coupling with the vector potential $\cal A$. Since  the axial-vector current carries positive electric charge $e$ (the elementary proton charge),  its  derivative is
\begin{equation}
    D\punto A(x) = [\partial_x-ie{\cal A}(x)]^\mu A_\mu(x).
\end{equation}
Hence, the new definition of the correlators in  configuration space is
\begin{align}
    \Pi^{AA}_{\mu\nu}(x,y) & = i\langle 0|T\, [ A_\mu(x)A^\dag_\nu(y)]|0\rangle\\
    \Pi_{5\nu}(x,y) & = i\langle 0|T\,[iD\punto A(x)\;A^\dag_\nu(y)]|0\rangle\\
    \psi_5(x,y) & = i\langle 0|T\,[iD\punto A(x)[i D\punto A(y)]^\dag] |0\rangle \,,
\end{align}
and the covariant derivative of the quark fields $q(x)$ becomes
\begin{equation}
    D_\mu\,q(x) \equiv [\partial_x-ie_q{\cal A}(x) - i\, g\, G(x)]_\mu \,q(x) \,.
\end{equation}
    
In the hadronic sector we use the axial-vector current field description from chiral perturbation theory ($\chi$PT) in terms of charged pion fields 
\begin{equation}
    A_\mu(x) = -f_\pi D_\mu\,\pi^+(x).
    \label{A-pion}
\end{equation}
In this case the covariant divergence of the axial-vector current is $D\punto A(x)=f_\pi\, m_\pi^2\,\pi^+(x)$.
This relation is obtained from the new equations of motion for the charged pion, $(D^2 +  M_\pi^2)\pi^+(x) = 0$, where the covariant derivative is defined as
\begin{equation}
    D_\mu\, \pi^+(x) = [\partial_x - ie{\cal A}(x)]_\mu\,\pi^+(x).
\end{equation}

\subsection{Ward identities}

The covariant derivative will modify the usual Ward identities in configuration space to the following
\begin{align}
    [i\partial_{x}+e{\cal A}(x)]^\mu \,\Pi^{AA}_{\mu\nu}(x,y) = 
    &\Pi_{5\nu}(x,y) -\Delta_\nu(x,y)\\
    [-i\partial_{y}+e{\cal A}(y)]^\nu \,\Pi_{5\nu}(x,y) =
    &\psi_5(x,y)+\Delta_5(x,y)
\end{align}
with
\begin{align}
    \Delta_\nu(x,y) & =\delta(x_0-y_0)\langle 0| [A_0(x),A^\dag_\nu(y)]|0\rangle\\
    \Delta_5(x,y) &=\delta(x_0-y_0)\langle 0|[D(x)\!\cdot\!\! A(x),A^\dag_0(y)]|0\rangle.
\end{align}
In the QCD sector, these $\Delta$ terms can be easily calculated through the quark anticommutation relations     
\begin{align}
    \Delta_\nu(x,y)\, &=\langle 0|(\bar d\gamma_\nu d -\bar u\gamma_\nu u)|0\rangle {(x)}\,\delta^4(x-y)\\
    \Delta_5(x,y)\,&= im_{ud}\langle 0|(\bar dd +\bar uu)|0\rangle {(x)}\,\delta^4(x-y),
\end{align}
and in the case of the hadron sector, using the commutation relation for pion fields gives  
\begin{align}
    \Delta_0(x,y) &=0\\
    \Delta_j(x,y) &=f_\pi^2[-i\partial_{y}+e{\cal A}(y)]_j\,\delta^4(x-y)\\
    \Delta_5(x,y)\, &=-i f_\pi^2m_\pi^2\,\delta^4(x-y)
\end{align}

In the presence of a magnetic field  the Schwinger phase  generates a non-locality in any current correlator, i.e., $\Pi(x,y) \neq \Pi(x-y)$.
Although this statement is in  general, our particular case is  not  affected by nonlocal terms. Hence we can define
\begin{equation}
    \Pi(q^2) \equiv \int d^4\!x \,e^{iq\cdot x}\,\Pi(x,0)
    = \int d^4\!y\, e^{-iq\cdot y}\,\Pi(0,y).
\end{equation}
The justification is as follows.
In the symmetric gauge the Schwinger phase vanishes if one of the coordinates is zero. 
Therefore, the phase factor vanishes if $x=0$ or $y=0$ in the case of one-loop PQCD diagrams,  tree-level diagrams (as in the hadronic sector), or diagrams  involving the chiral condensate.
The only diagrams under consideration that may include nonvanishing phase factors are the  gluon condensate ones.
Nevertheless, the phase factor can be expanded in powers of the magnetic field, while increasing the inverse power of momentum $1/q^{2N}$. 
We find that these contributions do not survive in the sum rules under consideration, since the gluon condensate is a dimension-four operator.
One should keep this in mind if higher dimensional sum rules are considered.
    
With the definition of the correlators in momentum space, the new Ward identities can be written as
\begin{align}
    Q^\mu\Pi_{\mu\nu}^{AA}(q^2) &= \Pi_{5\nu}(q^2)-\Delta_\nu(q^2)\\
    Q^{*\nu}\Pi_{5\nu}(q^2) &= \psi_5(q^2)+\Delta_5(q^2),
\end{align}
where $Q$ includes the vector potential of the external field in momentum space. 
Considering the symmetric gauge, the vector potential is ${\cal A}(x) = -\frac{1}{2}F_{\mu\nu}x^\nu$.
In this case the covariant derivative in momentum space is defined as
\begin{equation}
    Q_\mu = q_\mu +\frac{ie}{2}F_{\mu\nu}\frac{\partial}{\partial q_\nu}.
    \label{Q}
\end{equation}

\subsection{Tensor structures}
\label{subsec:tensor}

We consider an homogeneous external magnetic field along the $z$ axis. The electromagnetic field tensor  can then be written in the convenient form
$F_{\mu\nu}=B\epsilon_{\mu\nu}^\perp$, 
with the perpendicular antisymmetric tensor defined as $\epsilon^\perp_{\mu\nu}= g_{\mu 1}g_{\nu 2}-g_{\mu 2}g_{\nu 1}$.
This term will appear in all tensor structures and lead to the separation of vectors into parallel and perpendicular projections. 
Another term entering  this analysis is the contraction of the external momentum with the antisymmetric perpendicular tensor $\tilde q_\mu \equiv \epsilon_{\mu\nu}^\perp q^\nu$.
The  metric is $g_{\mu\nu}=g_{\mu\nu}^\parallel +g_{\mu\nu}^\perp$, so that e.g. $q_\perp^2 = -\boldsymbol{q}_\perp^2$.
The magnetic field introduces several modifications in the tensor structure of the current correlators. 
Basically it consists of any combination of $q_\mu$ $g_{\mu\nu}$ and $\epsilon_{\mu\nu}^\perp$ which produces a rich variety of new independent components, usually associated with new condensates.
For instance, for $\Pi_{\mu\nu}^{AA}(q^2)$  the possible structures are $g_{\mu\nu}$, $\epsilon_{\mu\nu}^{\perp}$, $g_{\mu\nu}^\perp$, and all the pair combinations of $q_\mu$, $q_\mu^\perp$ and $\tilde q_\mu$.
Similarly, the possible structures for $\Pi_{5\nu}(q^2)$ are $q_\mu$, $q_\mu^\perp$, and $\tilde q_\mu$.

A simple way to isolate a given contribution is to project  it such as to exclude all other possibilities.
For instance, the $\Pi_0(q^2)$ function of the axial-vector correlator can be obtained  as
\begin{equation}
    \Pi_0(q^2) = \left(2\frac{q_\parallel^\mu q_\parallel^\nu}{q_\parallel^4}-\frac{g_\parallel^{\mu\nu}}{q_\parallel^2}\right)\Pi_{\mu\nu}^{AA}(q^2)\,.
\end{equation}
Similarly the $\Pi_5(q^2)$ component of $\Pi_{5\nu}(q^2)$,
not the only term in a magnetic field,
  can be obtained as
\begin{equation}
    \Pi_5(q^2)=\frac{q_\parallel^\mu}{q_\parallel^2}\Pi_{5\nu}(q^2)\,.
\end{equation}
\begin{figure}[ht]
	\begin{center}
		\includegraphics[height=.17\textheight, width=.26\textwidth]{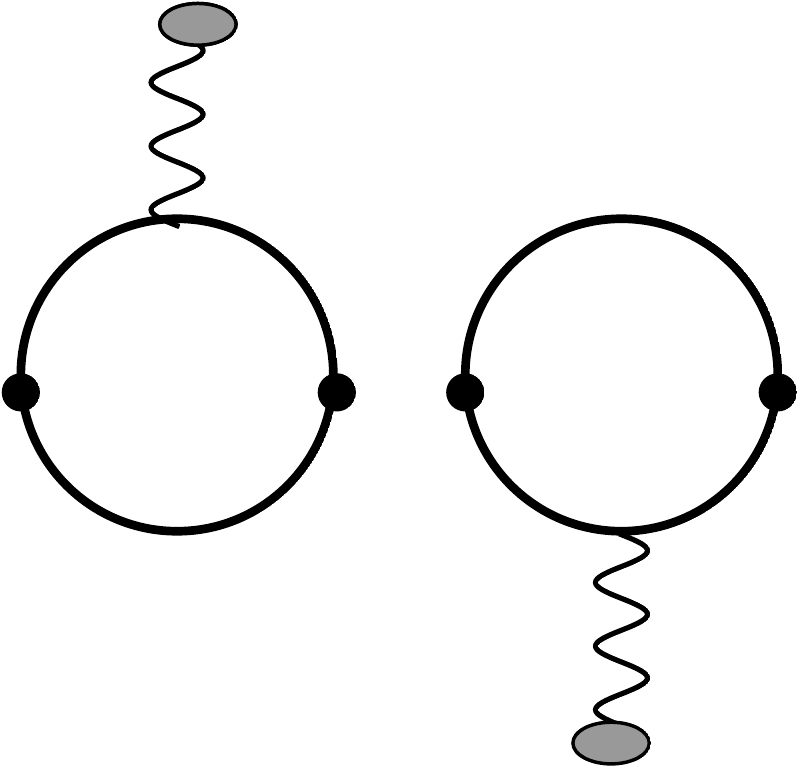}
		\caption{{\small QCD current correlator to leading order in the magnetic field, ${\cal{O}}(eB)$. Upper  line is the up-quark, and lower line the down-quark. Wiggly line represents the interaction with the external magnetic field.}}     
		\label{fig:OeB1}
	\end{center}
\end{figure}
  
\section{Magnetic field contribution to the current correlators}
 
The presence of a magnetic field is to be reflected in the charged particle propagators. These will be expressed in a power series involving the magnetic field \cite{Chyi:1999fc}.  The quark  and the pion propagator become 
\begin{align}
    S_q(x,y)=e^{ie_q\phi(x,y)}\int \frac{d^4 k}{(2\pi)^4}e^{-ik\punto (x-y)}\sum_n S_q^{(n)}(k)\\
    D_\pi(x,y)=e^{ie_\pi\phi(x,y)}\int \frac{d^4 p}{(2\pi)^4}e^{-ik\punto (x-y)}\sum_n D_\pi^{(n)}(p)\,,
\end{align}
with $e_{i}\phi$ the Schwinger phase,  $e_{i}$ the corresponding particle-charge,   $\phi$ defined in the symmetric gauge as
\begin{equation}\label{phase}
    \phi(x,y)=-\frac{1}{2}F_{\mu\nu}x^\mu y^\nu\,,
\end{equation}
 and the index $n$ in the sums referring to the power in the field, i.e. $B^n$.
It is important to point out that the series is well defined along the contour in the complex s-plane, Fig.~\ref{fig:figure1}. This is due to the integration path not crossing through the positive real $s$ axis, except  at $s_0$ generating the  discontinuity. Hence,  the only terms needed for  magnetic corrections in QCD are the following
\begin{align}
    S_q^{(0)}(k) = &~ i\frac{\slashed{k}+m_q}{k^2-m_q^2}\\
    S_q^{(1)}(k) = &~ -\gamma_1\gamma_2 (e_q B) \frac{(\slashed{k}_{\parallel} + m_q)}{(k^2 - m_q^2)^2}\\
    S_q^{(2)}(k) = &~  \frac{2 i\, (e_q B)^2}{(k^2 - m_q^2)^4}  \left[ k_\perp^2 (\slashed{k} + m_q) - \slashed{k}_\perp ( k_{\parallel}^2 - m_q^2) \right].
\end{align}

There will be infrared divergences from the magnetic contributions, which are safely controlled by the magnetic quark masses. Hence, it is necessary
to keep finite quark masses to leading order  in expansions in terms of $m_q/s_0$. \\
In the case of the pion,  the only contribution is that of the  propagator at zero magnetic field $D_\pi^{(0)}(p)=i/(p^2-m_\pi^2)$. This is because the next term is $D_\pi^{(1)} = 0$,  and the other terms do not survive in the FESR under consideration.

\subsection{PQCD sector}  
\begin{figure}[ht]
\includegraphics[height=.27\textheight, width=.25\textwidth]{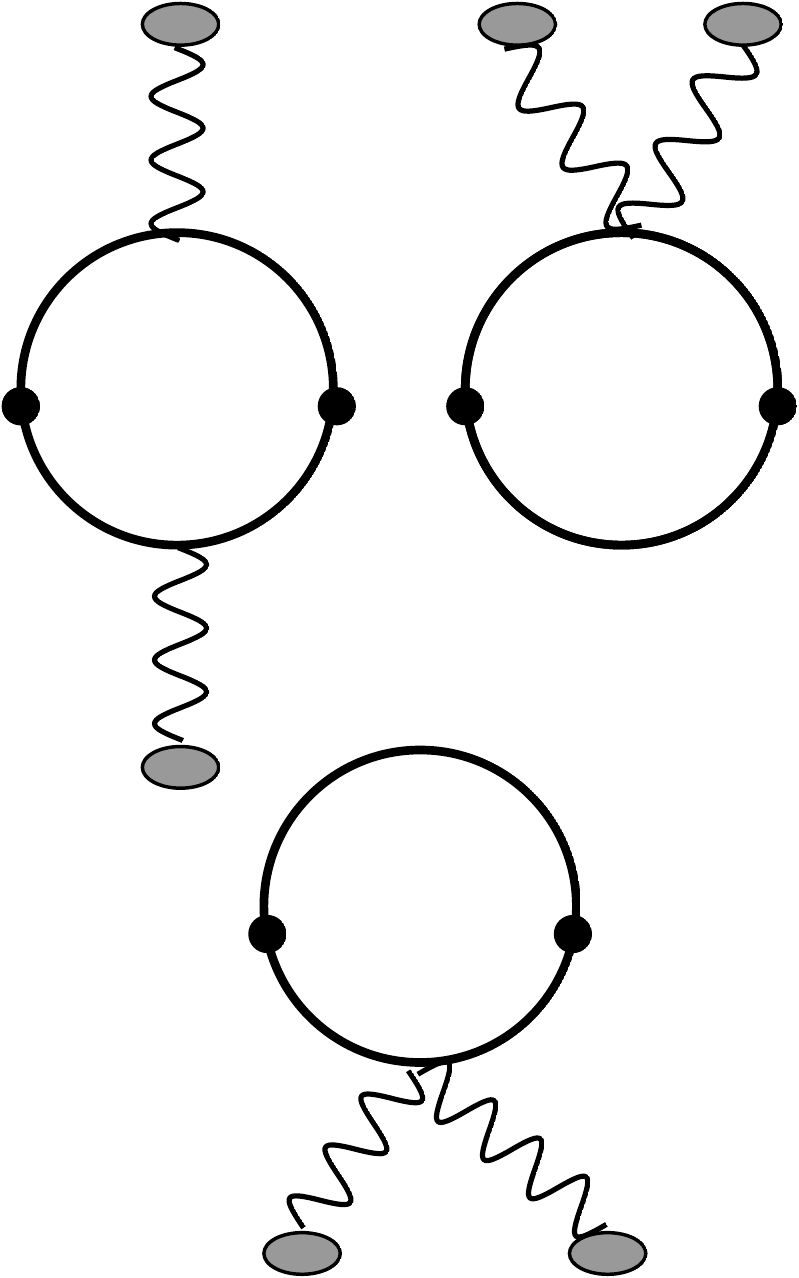}
\caption{{\small QCD current correlator at next-to-leading order in the magnetic field, ${\cal{O}}[(eB)^2]$. Upper line is the up-quark, and lower line the down-quark. Wiggly line represents the interaction with the external magnetic field. The current correlator associated with the first diagram is labeled as  $\Pi_{\mu\nu}^{(1,1)}$, the second as  $\Pi_{\mu\nu}^{(2,0)}$, and the third as $\Pi_{\mu\nu}^{0,2}$.}}    
\label{fig:OeB2}
\end{figure}

The leading order magnetic field correction to a current correlator in QCD, $ {\cal{O}} (eB)$, is indicated in Fig.~\ref{fig:OeB1} as a wiggly line attached to the up-quark and to the down quark (upper/lower solid lines, respectively). The next-to-leading order contribution, ${\cal{O}} (eB)^2$, is shown in Fig.~\ref{fig:OeB2}
No additional contributions are needed, as they do not contribute to the chosen FESR.
As  mentioned earlier, the Schwinger phase in one-loop PQCD diagrams vanishes in the symmetric gauge, after setting the coordinate $y=0$ or $x=0$.
      
Starting with the axial-vector current correlator,
the structures in the PQCD sector  contributing to the two FESR are
\begin{multline}\label{PiAA_structure}
    \Pi_{\mu\nu}^{AA}(q^2) 
        = q_\mu q_\nu \,\Pi_0 (q^2) 
        + g_{\mu\nu}\,\Pi_1(q^2) 
         + i\epsilon_{\mu\nu}^\perp  \,\tilde\Pi_1(q^2) \\
        + q^\perp_\mu q^\perp_\nu \,\Pi_0^\perp (q^2)
        + g_{\mu\nu}^\perp \,\Pi_1^\perp (q^2) \\
        +(q_\mu q_\nu^\perp+q_\nu q_\mu^\perp)\,\Pi_2(q^2)
\end{multline}
Notice that  only  $\Pi_0(q^2)$ is relevant.

The diagrams contributing to $\Pi_0(q^2)$ are shown in Fig.~\ref{fig:OeB2}. 
The diagrams of Fig.~\ref{fig:OeB1} contribute only to $\tilde\Pi_1(q^2)$, entering Eq.~(\ref{PiAA_structure}).
The next-to-leading order magnetic field correction involves three diagrams, as  shown in Fig.~\ref{fig:OeB2}. They are labeled  (1,1), (2,0), and (0,2), respectively, denoting the order of magnetic insertions in $(u,d)$ . 
Unlike the leading order magnetic correction, and the (1,1) term, the diagrams (2,0) and (0,2) are infrared divergent. Hence, quark masses must remain nonzero. \\
With the frame choice $q_{\perp} =0$,  and $q_{\parallel}^2 \equiv s$, the relevant magnetic contributions to $\Pi_0$ are

\begin{align}
    \Pi_0^{(1,1)}(s) = &\,\frac{3}{2\, \pi^2} e_ue_dB^2\int_0^1 dx\,\frac{1}{[s - M^2(x)]^2}\,,\label{Pi011}\\
    \Pi_0^{(2,0)}(s) = &\,\frac{1}{2 \pi^2} \, (e_uB)^2\, \int_0^1 \frac{dx}{1-x}\,\frac{x^2}{[s - M^2(x)]^2}\,, \label{Pi020}\\
    \Pi_0^{(0,2)}(s) = &\, \frac{1}{2 \pi^2} \, (e_d B)^2\, \int_0^1 \frac{dx}{x}\,\frac{(1 - x)^2}{[s - M^2(x)]^2}\,, \label{Pi002}
\end{align}
where the quark charges are defined as
\begin{equation}
    e_u=\frac{2}{3}e,
    \qquad
    e_d=-\frac{1}{3}e,
\end{equation}
and where
\begin{equation}
M^2(x) \equiv \frac{m_u^2}{1 - x} \, +\, \frac{m_d^2}{x}\,.\label{M2}
\end{equation} 

It should be noticed from the results for the second and third diagrams in Fig.~\ref{fig:OeB2}, Eq.(\ref{Pi020}) and (\ref{Pi002}), that logarithmic quark-mass (infrared) singularities will appear as a consequence of  magnetic field overlapping. In QCD in the vacuum, logarithmic light-quark mass singularities in current correlators appear at next to leading order in perturbation theory. They can be removed  by a suitable procedure \cite{IRmq1,IRmq2}. The situation here is rather different in that the source of the singularities is the presence of the external magnetic field, at leading order in perturbative QCD.

The general structure of $\Pi_{5\nu}$ is
\begin{equation}\label{Pi5_structure}
\Pi_{5\nu}(q^2) = q_\nu\Pi_5(q^2)+q^\perp_\nu\Pi_5^\perp(q^2)+i\tilde q_\nu\tilde\Pi_5(q^2)
\end{equation}
where we are interested only in $\Pi_5(q^2)$. 
The diagrams in Fig.~\ref{fig:OeB1} only contribute to $\tilde\Pi_5(q^2)$ , as indicated in Eq.~(\ref{Pi5_structure}).
The magnetic contributions to $\Pi_5(q^2)$,  are of  order $(s-M^2)^{-2}$. Therefore they do not contribute to the FESR under consideration ($P(s)=1$). 
This fact will be discussed in more detail in the next section.

Finally, the correlator involving the axial-vector current divergences   has only one structure and the magnetic contributions arise from the diagrams  in Fig.~\ref{fig:OeB2}.
Choosing $q_\perp=0$ and $q_\parallel^2\equiv s$, gives
\begin{align}
    \psi_5^{(1,1)}(s) &=\, -\frac{3}{4\pi^2}{m_{ud}}^2\,e_u e_d B^2
        \int_0^1 dx\,\frac{1}{s-M^2(x)} \label{Psi511}\\
    \psi_5^{(2,0)}(s) &=\, \frac{1}{4\pi^2}{m_{ud}}^2\,(e_u B)^2
        \int_0^1 \frac{dx}{1-x}\,\frac{x}{s-M^2(x)} \label{Psi520}\\
    \psi_5^{(0,2)}(s) &=\, \frac{1}{4\pi^2}{m_{ud}}^2\,(e_d B)^2
        \int_0^1 \frac{dx}{x}\,\frac{1-x}{s-M^2(x)},\label{Psi502} 
\end{align}
where   terms of order $(s-M^2)^{-2}$ are omitted as they do not contribute to the FESR  with $P(s)=1$, unlike the case of $\Pi_5(q^2)$.
Notice  that logarithmic terms are also   present in Eq. (\ref{Psi520}) and (\ref{Psi502}).

While the integration in the variable $x$ is rather complicated, ultimately these expressions  enter the contour integral in the complex s-plane. 
This feature simplifies considerably the integration, as discussed in the next section.

\subsection{The nonperturbative QCD sector}
\label{subsec:NPQCD}

In the nonperturbative QCD sector both the quark  and the gluon condensates develop a magnetic field dependence. They will be determined by the FESR themselves. Regarding the quark condensate in the presence of a magnetic field there is an additional contribution from a condensate  $\langle \bar{q} \, \sigma_{12} \, q \rangle$, where $\sigma_{\mu\nu}=i[\gamma_\mu,\gamma_\nu]/2$ \cite{Ioffe:1983ju} (see also Gatto and Ruggieri in \cite{Kharzeev:2013jha}, and references therein).
This condensate does not appear in $\Pi_0(q^2)$ used in the FESR with $P(s)=1$ and $P(s)=s$, nor in $\Pi_5(q^2)$ or $\psi_5(q^2)$ for $P(s)=1$. 
However this term will be present in $\Pi_5(s)$ with kernel $P(s)=s$.
The fact that $\Pi_5(s)$ with $P(s)=s$ and $\psi_5(s)$ with $P(s)=1$ provide different information at finite $eB$, unlike the vacuum case, is related to the new Ward identities. 
The new condensate $\langle\bar q\sigma_{12}q\rangle$ can be calculated from FESR using other structures. Although it is an interesting  contribution, it is beyond the scope of this analysis.
Another issue to be considered is that in a magnetic field the quark condensates and the quark masses will be  flavor dependent. Hence,  
the contributions to the chiral condensate in Sec.~\ref{sec:vacCorr} change as follows: $\langle\bar qq\rangle\to \langle\bar uu+\bar dd\rangle(1-\Delta_{ud})/2$ in Eq. (\ref{Pi0AANP}), and  $\langle\bar qq\rangle\to \langle\bar uu+\bar dd\rangle/2$ in Eq. (\ref{Pi5QCD}), and $\langle\bar qq\rangle\to \langle\bar uu+\bar dd\rangle(1-3\Delta_{ud})/2$ in Eq. (\ref{Psi5QCD}), where 
\begin{equation}
    \Delta_{ud}\equiv \frac{m_u-m_d}{m_u+m_d}\,\frac{\langle\bar uu-\bar dd\rangle}{\langle\bar uu+\bar dd\rangle}.
\end{equation}
An estimate of  $\Delta_{ud}$ can be obtained by considering $m_d\approx 2m_u$ and including the values of the condensates obtained  at finite magnetic field in \cite{Bali:2012zg}. This gives $\Delta_{ud} \sim 0.1$. 
In particular, for $eB=0.2\text{ GeV}^2$ one has $\Delta_{ud}=0.013$, and for $eB=1\text{ GeV}^2$ one has $\Delta_{ud}=0.067$.
Hence, one can safely ignore this correction. 
        
\bigskip

\begin{figure}
    \includegraphics[scale=0.8]{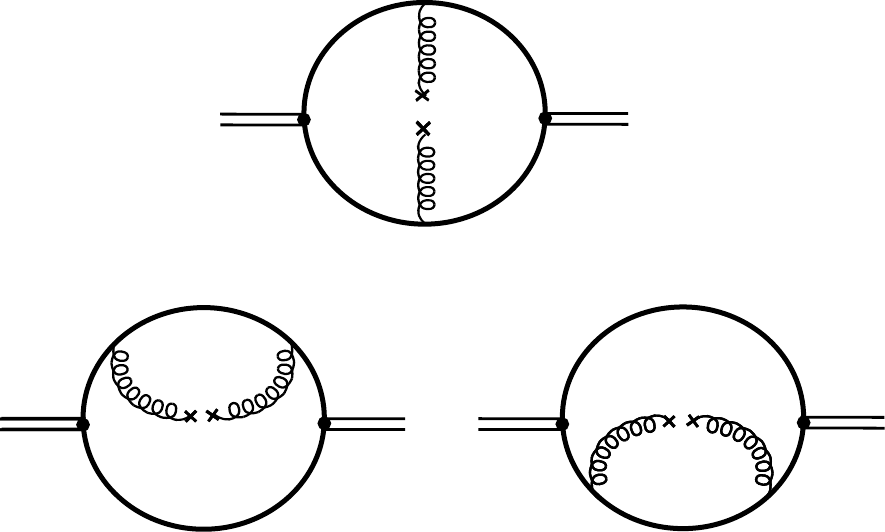}
    \caption{\small Contribution of the gluon condensates to a two-point correlator.}
    \label{fig:GG}
\end{figure}

The case of the gluon condensate must be treated with care. The diagrams involving the gluon condensate have several Schwinger phase terms, with not all depending on $x,y$. Hence, they do not vanish for our choice of  gauge. In detail,
in the symmetric gauge the Schwinger phase can be written as in Eq. (\ref{phase}). The diagrams  contributing to the gluon condensate are shown in Fig.~\ref{fig:GG}. 
Considering e.g. the bottom  left  diagram   in Fig.(\ref{fig:GG}), omitting $\gamma$-matrices it
can be written as 
\begin{equation}
    \Pi^{\langle\alpha G^2\rangle}(x,y)\propto \text{tr}\int_{zw}S_u(x,z)S_u(z,w)S_u(w,y)S_d(y,x).
\end{equation}
The phase of the propagator at the centre of this  equation does not vanish if $x=0$ or $y=0$. 
However, one can expand these phases as
\begin{equation}
    e^{ie_u\phi(z,w)}=1-i \frac{e_u}{2}F_{\mu\nu}z^\mu w^\nu +\dots
\end{equation}
The powers in coordinates correspond to derivatives in momentum space which will increase the power of the denominator in the propagator. 
Hence, as the correlators involving the gluon condensate for $\phi=0$ are of order $(s-M^2)^{-2}$, the next contribution to the phase expansion generates a term $\sim (s-M^2)^{-3}$, or  a  higher order denominator. All these terms
vanish in the FESR under consideration.

\bigskip

Finally,  in principle there are new condensates due to Lorentz symmetry breaking in an external magnetic field.
For instance, the gluon condensate term $\langle \alpha_s {G_{\mu\nu}}^2\rangle$ should split their components into parallel and perpendicular, or mixed contributions \cite{DElia:2015eey}. 
In addition,
 the condensate $\langle \bar q\slashed{D}q\rangle = -im_q\langle\bar qq\rangle$ should split into $\langle \bar q\slashed{D}_\parallel q\rangle$
and $\langle \bar q\slashed{D}_\perp q\rangle$.
This kind of splitting is associated with the new tensor structure mentioned in Sec. \ref{subsec:tensor}.
The role of such objects will be explored elsewhere.

\subsection{Hadronic sector}

Next, we  consider the hadronic contribution to the current correlators. The pion contribution to the axial-vector current is given in  Eq.~(\ref{A-pion}). The axial-vector correlator in momentum space is
\begin{equation}
    \Pi_{\mu\nu}^{AA}(q^2)|_\text{\tiny Had} = 
   i 2f_\pi^2Q_\mu Q^*_\nu  D_\pi(q)+ 2f_\pi^2g_{\mu 0}g_{\nu 0}
\end{equation}
where $Q$ is defined in Eq.~(\ref{Q}).
The constant term on the right-hand side (r.h.s) does not contribute to the sum-rules. Ignoring it,  the axial-vector current correlator becomes
\begin{multline}\label{PiAA_structure-had}
    \Pi_{\mu\nu}^{AA}(q^2) 
        = q_\mu q_\nu \,\Pi_0 (q^2) 
         + i\epsilon_{\mu\nu}^\perp  \,\tilde\Pi_1(q^2) 
        + g_{\mu\nu}^\perp \,\Pi_1^\perp (q^2) \\
        + \tilde q_\mu \tilde q_\nu \,\tilde\Pi_0 (q^2) 
        +i(q_\mu \tilde q_\nu-q_\nu \tilde q_\mu)\,\Pi_3(q^2)
\end{multline}

Similarly, using the equation of motion for the charged pion,  $\Pi_{5\nu}(q^2)$ and $\psi_5(q^2)$ become
\begin{equation}
    \Pi_{5\nu}(q^2)|_\text{\tiny Had} = 2 \,i\,\pi^2\, m_\pi^2\,Q^*_\nu \,D_\pi(q^2)
\end{equation}
and
\begin{equation}
    \psi_{5}(q^2)|_\text{\tiny Had} =  2 \,i\, f_\pi^2\, m_\pi^4\, D_\pi(q^2),
\end{equation}
respectively.
Finally,   $\Pi_{5\nu}$ is given by
\begin{equation}
    \Pi_{5\nu}(q^2)|_\text{\tiny Had} =
    q_\nu\Pi_5(q^2)+i\,\tilde q_\nu\tilde\Pi_5(q^2).
\end{equation}

As mentioned earlier, the next to leading order in the magnetic field expansion in powers   of the pion propagator is proportional to $(s-m_\pi^2)^{-3}$. Hence, it does not contribute to the FESR under consideration. Thus,  the correlators $\Pi_0(q^2)$, $\Pi_5(q^2)$ and $\Pi_{5\nu}(q^2)$ involve the same hadronic spectral functions given in Eq. (\ref{ImPi0}), (\ref{ImPi5}) and (\ref{ImPsi5}).

\section{QCD finite energy sum rules in an external magnetic field}

We consider first the contour integrals involving the overlapping magnetic field contributions, as given  in Eq. (\ref{Pi020}), (\ref{Pi002}), (\ref{Psi520}) and (\ref{Psi502}). It is important to notice that in the vacuum, and  even  in  the chiral limit there will be magnetic mass generation. Before integrating in the Feynman parameter it is more convenient to first integrate in the momentum.
The magnetic contribution to the contour integral in the complex squared-energy s-plane is given by
\begin{equation}
I^N_n(s_0) = \frac{-1}{2 \pi i} \oint_{\text{C}(s_0)} ds \, s^{N-1} \int_0^1dx\,\frac{f(x)}{[s - M^2(x)]^n}, \label{InNs_0}
\end{equation}

where $M^2$ is defined in Eq.(\ref{M2}), and $f(x)$ is an arbitrary   function of $x$. In particular, this contribution is infrared divergent for massless quarks as $f\propto 1/x$ and $f\propto 1/(1-x)$.
If $N< n$, this integral vanishes, while
if $N \geq n$ this integral is nonvanishing only if $M^2 (x) < s_0$.
The nonvanishing FESR considered here are for $N=n$, so that after integrating in $s$,  $I^N_n(s_0)$ becomes 
\begin{equation}
I^n_n(s_0)=
-\int_0^1dx \,f(x)\, \theta\left(s_0 - M^2(x)\right)\,. 
\label{Inn}
\end{equation}

The restriction imposed by the $\theta$- function leads to the  quadratic equation 
\begin{equation}
    (x-x_+)(x-x_-)<0, 
    \label{dicontinuity}
\end{equation}
with $x_+>x_-$,  and $x_\pm$ defined as
\begin{multline}
    x_\pm=\frac{1}{2}\Bigg[1+\frac{m_d^2}{s_0}-\frac{m_u^2}{s_0}\\
    \left.\pm \sqrt{1-2\left(\frac{m_d^2}{s_0}+\frac{m_u^2}{s_0}\right)+\left(\frac{m_d^2}{s_0}-\frac{m_u^2}{s_0}\right)^2 }\,\right].
    \label{xpm}
\end{multline}
The  inequality, Eq. (\ref{dicontinuity}), is satisfied only for $x_-<x<x_+$, so that Eq. (\ref{Inn}) can be rewritten as
\begin{equation}
    I^n_n(s_0)=
    -\int_{x_-}^{x_+}dx \,f(x).
\end{equation}

The series expansion will be carried out for $m_q^2\ll s_0$ and up to first order.
Hence, the integration limits in Eq. (\ref{xpm}) can be approximated as
\begin{align}
    x_+ &\simeq 1-m_u^2/s_0,\\
    x_- &\simeq m_d^2/s_0,
\end{align}
which allows  handling the IR divergences.
After the expansion in quark masses, there will appear flavor dependent logarithmic terms.  
After separating the average part from the mass difference part, these terms become
\begin{align}
    \ln(s_0/m_u^2) = &\ln(4s_0/{m_{ud}}^2)-2\ln(1+\delta_{m})\\
    \ln(s_0/m_d^2) = &\ln(4s_0/{m_{ud}}^2)-2\ln(1-\delta_{m})
\end{align}
with $\delta_m \equiv (m_u-m_d)/(m_u+m_d)$. 
The contribution of $\delta_m$  is negligible compared to the logarithm term. 
A numerical estimate for $m_d=2m_u\simeq 10\text{ MeV}$, and $s_0\simeq 1\text{ GeV}^2$ shows that $2\ln(1\pm\delta_m)$ is at least one order of magnitude smaller than $\ln(4s_0/{m_{ud}}^2)$.
Therefore, the mass-difference contribution can be safely neglected. 

\section{Results}

The FESR involving magnetic field corrections are
\begin{align}
    2f_\pi^2 
            =&\; 
        \frac{s_0}{4\pi^2}  +{\cal O}({m_{ud}}^2) \\
            &\nonumber\\
    2f_\pi^2m_\pi^2 
        =&\;
        \frac{1}{8\pi^2}\left\{s_0^2
        -\frac{2}{9}(eB)^2\left[10\ln(4s_0/{m_{ud}}^2)-27\right]\right\}
            \nonumber\\ &
        +\frac{1}{2}m_{ud}\langle\bar uu+\bar dd\rangle
        -\frac{1}{12\pi}\langle\alpha_s G^2\rangle 
        +{\cal O}({m_{ud}}^2 s_0)\\
            &\nonumber\\
    \frac{2f_\pi^2m_\pi^2}{m_{ud}}
        =&\;
        - \langle\bar uu+\bar dd\rangle 
        +\frac{3}{8\pi^2}{m_{ud}} s_0
        +{\cal O}({m_{ud}}^3)\\ \label{SR3-eB}
            &\nonumber\\
    \frac{2f_\pi^2m_\pi^4}{{m_{ud}}^2}
  =&\;
    \frac{3}{16\pi^2}\left\{s_0^2-\frac{20}{27} (eB)^2\left[
    \ln(4s_0/{m_{ud}}^2)-1
    \right]\right   \}
        \nonumber\\&
    -\frac{1}{4} m_{ud}\langle\bar uu+\bar dd\rangle
    +\frac{1}{8\pi}\langle\alpha_sG^2\rangle
        +{\cal O}({m_{ud}}^2s_0).
 \end{align}
where $m_\pi$ and $f_\pi$ are functions of the magnetic field. The only restriction  is  ${m_{ud}}^2\ll s_0$, which remains valid for all  values of $eB$ under consideration. 
There are six parameters to be determined, i.e. 
$m_{ud}$, $m_\pi$, $f_\pi$, $s_0$, $\langle \bar qq\rangle$, and $\langle\alpha_sG^2\rangle$. 
Since there are  only four independent FESR,  two inputs are required. 
We separate them into vacuum inputs and magnetic evolution inputs:
\begin{enumerate}
\item
The vacuum parameters are the charged pion mass and the pion decay constant. 
\item
 As a first  input, we choose the magnetic evolution of the quark condensates from NJL results \cite{Coppola:2018vkw}, which agree with LQCD \cite{Bali:2012zg}.
\item
For the second magnetic input we  choose three different  scenarios:
\begin{enumerate}
\item
The first scenario involves the magnetic evolution of the charged pion mass provided by NJL calculations \cite{Coppola:2018vkw}. 
\item
The second one involves the linear relation between $m_\pi^2$ and $m_{ud}$, i.e. $m_\pi^2 ={\cal {B}}\, m_{ud}$ from the Nambu-Goldstone realization of chiral $SU(2)\times SU(2)$ symmetry. 
It is assumed that $\cal {B}$ is independent of the magnetic field, i.e.  $m_{ud}/m_\pi^2=$  constant.
\item
In the third scenario,  the quark masses are assumed magnetic field independent, i.e $m_{ud}=$ constant.
\end{enumerate}
\end{enumerate}

In principle one could assume that $f_\pi^2$ depends on the magnetic field as the quark condensate does. However, this leads to unexpected results, e.g.  negative quark mass values, implying a vanishing pion mass. An  interesting consequence of the  magnetic dependence of quark masses is that  the GMOR relation either breaks down, or is modified as
\begin{equation}
    m_\pi^2f_\pi^2 = \frac{m_{ud}}{1+\frac{3}{2}\frac{{m_{ud}}^2}{m_\pi^2}}\langle \bar uu +\bar dd \rangle.
\end{equation}

This kind of modification was obtained in \cite{Simonov:2012ym, Orlovsky:2013wjd}, where magnetic dependent  quark masses were considered.\\

\begin{figure}
\centering
\includegraphics[scale=0.6]{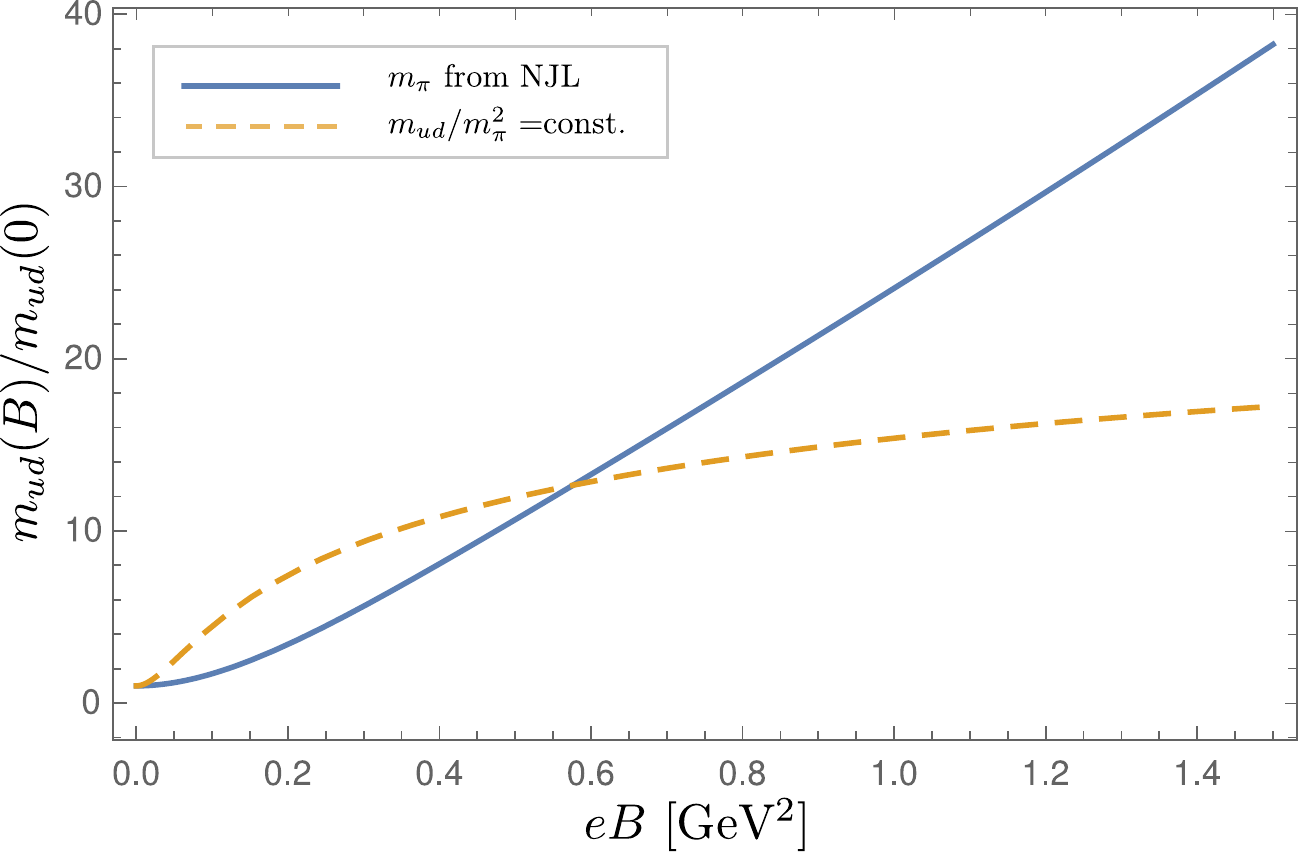}
\caption{\small Magnetic evolution of the sum of the up- and down-quark masses for two of the conditions used as an input.}
\label{plot:mud}
\end{figure}

Figure~\ref{plot:mud} shows the magnetic evolution of the normalized quark mass $m_{ud}$. For high values of $eB$ it increases approaching  the constituent quark mass. A similar effect was found for the thermal evolution of $m_{ud}$ \cite{qmassT}.

\begin{figure}
\centering
\includegraphics[scale=0.6]{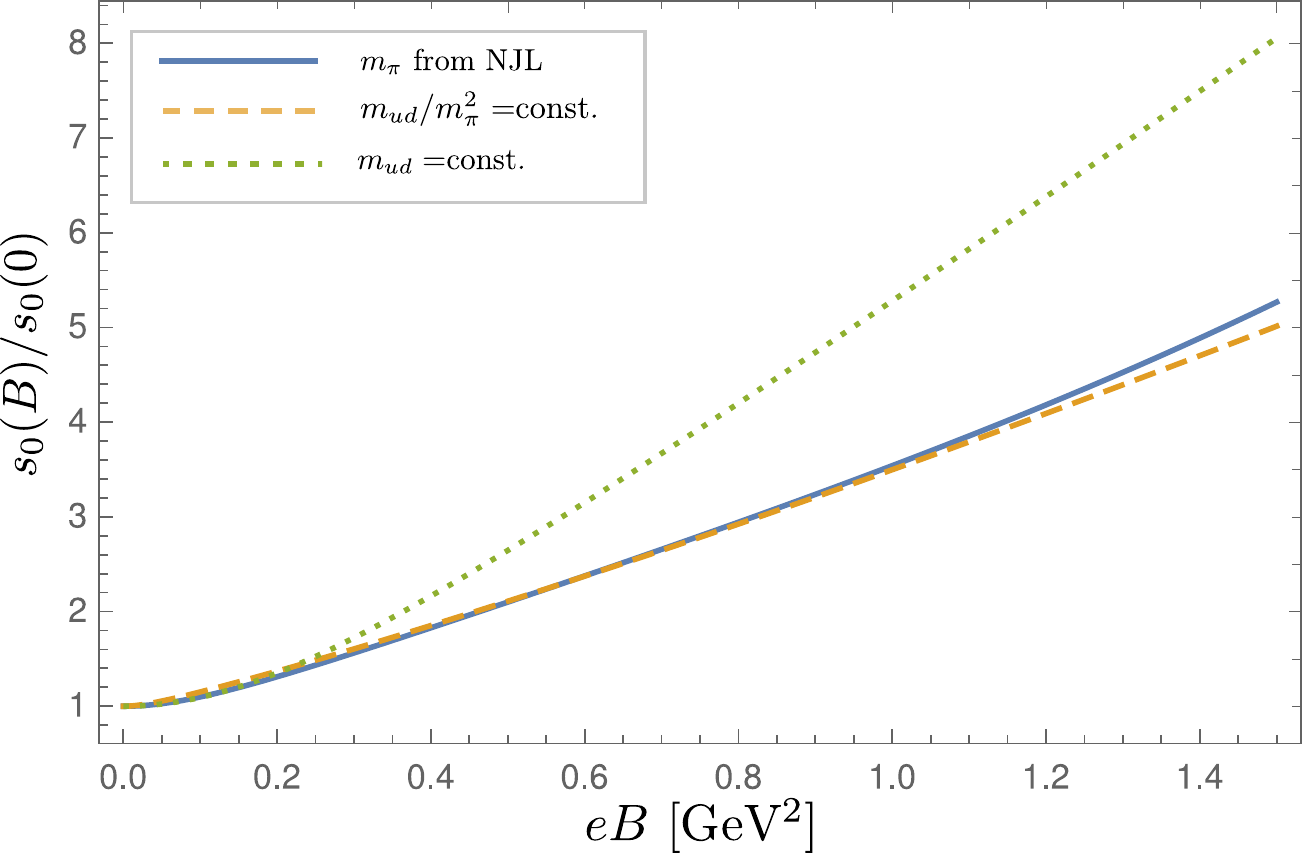}
\caption{\small  Magnetic evolution of the continuum hadronic threshold for the three conditions used as an input.}
\label{plot:s0}
\end{figure}

\begin{figure}
\centering
\includegraphics[scale=0.6]{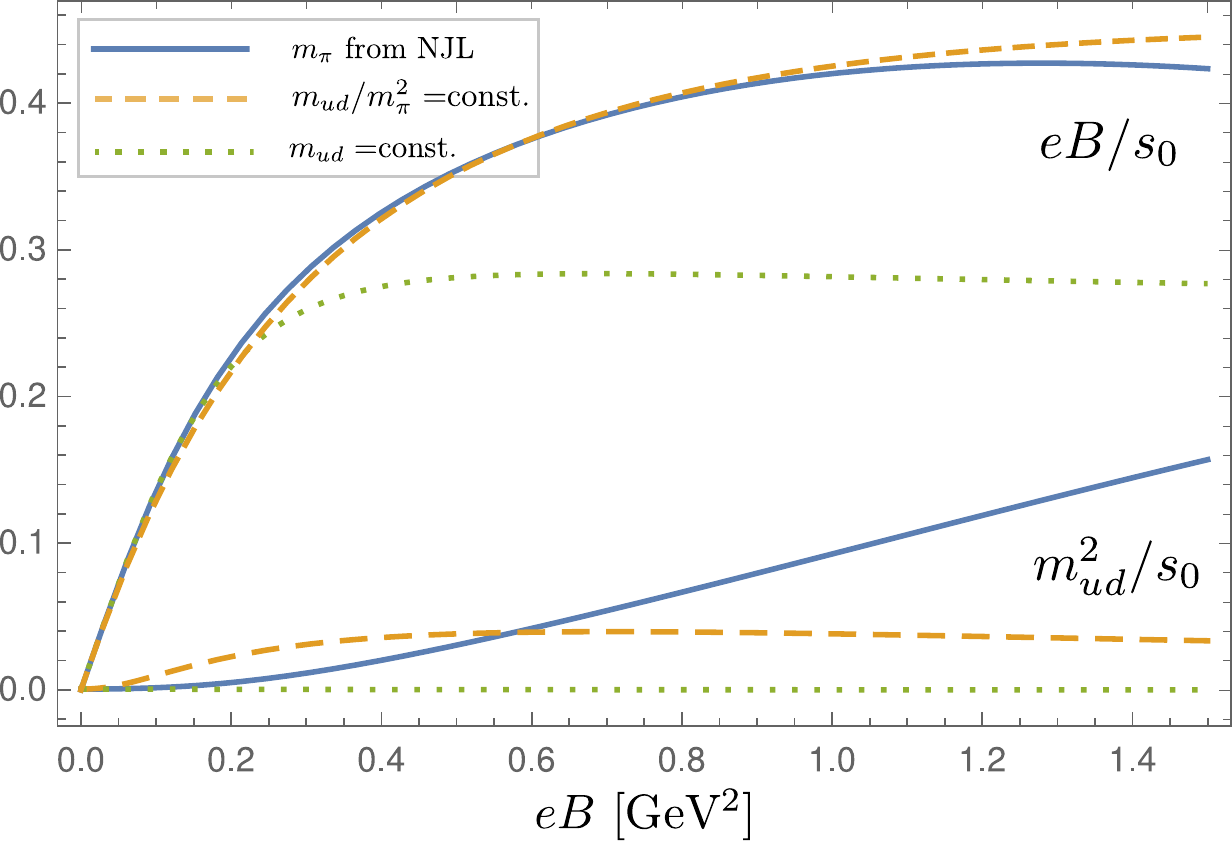}
\caption{\small 
Evolution of the ratios ${m_{ud}}^2/s_0$ and $eB/s_0$ considering the three conditions used as an input.}
\label{plot:mud^2_s0}
\end{figure}

The magnetic evolution of  $s_0$ is shown in Fig.~\ref{plot:s0} for the three input schemes. 
This behavior validates the relation  ${m_{ud}}^2\ll s_0$ as   seen in Fig.~\ref{plot:mud^2_s0}. 
Also,  the ratio $eB/s_0$ in Fig.~\ref{plot:mud^2_s0} shows  $s_0$ to be always the dominant scale in this range. 

\begin{figure}
\centering
\includegraphics[scale=0.6]{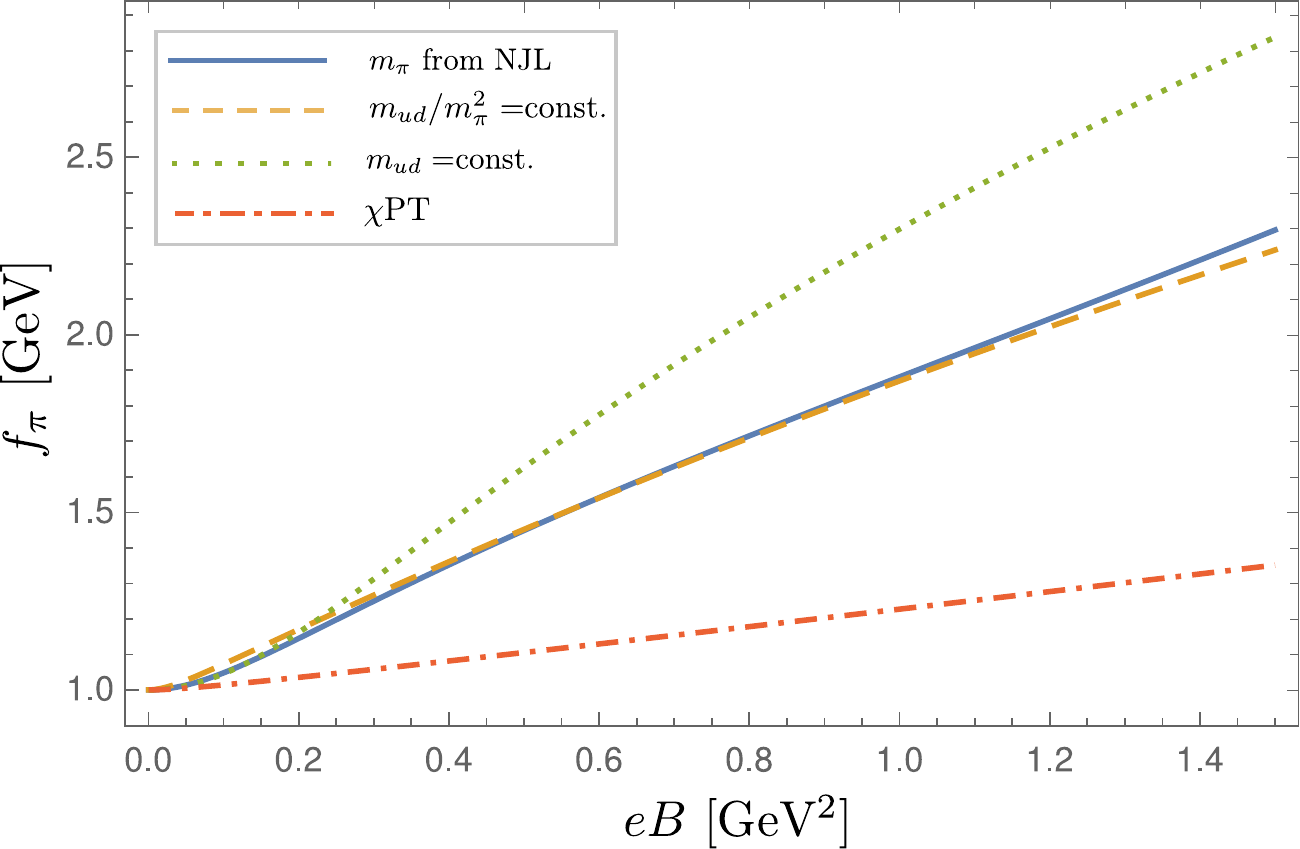}
\caption{\small  Magnetic evolution of $f_\pi$ considering the three conditions used as an input.
For comparison, we include the $\chi$PT result from Ref. \cite{Andersen:2012zc}.
}
\label{plot:fpi}
\end{figure}

Figure \ref{plot:fpi} shows the magnetic evolution of the pion decay constant for both quark mass schemes, as well as the case using  results from $\chi$PT \cite{Andersen:2012zc}.
Notice that the input using the NJL pion mass and the input for  $m_{ud}/{m_\pi}^2$ generate a similar behavior of $f_\pi$. 
This is perhaps the most robust prediction of this analysis.\\

We recall that the magnetic evolution of $s_0$ and  $f_\pi^2$ are identical
\begin{equation}
\frac{s_0(B)}{s_0(0)} = \frac{f_\pi^2(B)}{f_\pi^2(0)}.
\end{equation}
However, the only scheme that leads to the same magnetic evolution of $f_\pi^2$ and the chiral condensate is the one considering a constant quark mass. \\

\begin{figure}
\centering
\includegraphics[scale=0.6]{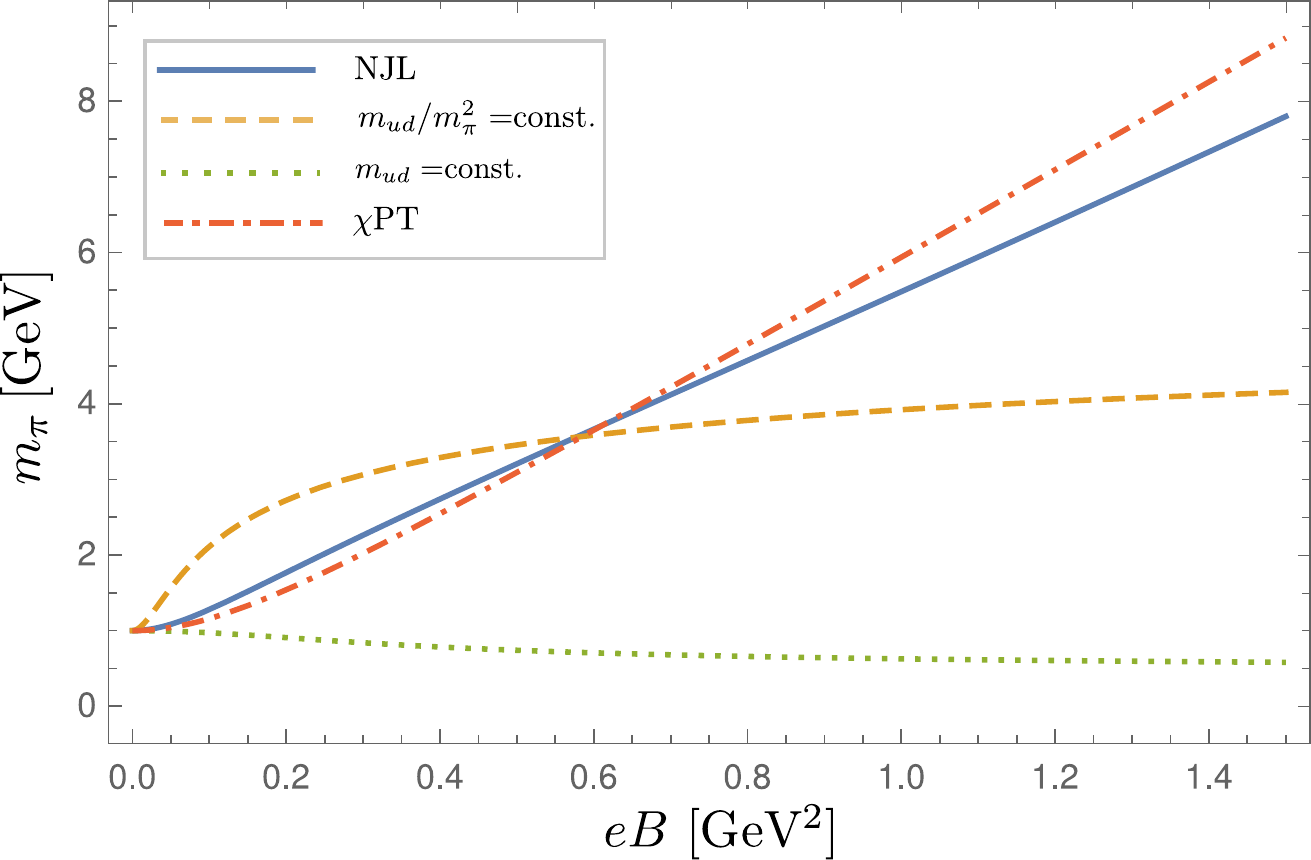}
\caption{\small Magnetic evolution of the charged pion mass. 
For comparison, we include the $\chi$PT result from Ref. \cite{Andersen:2012zc}.}  
\label{plot:mpi}
\end{figure}

In Fig.\,\ref{plot:mpi} we show the magnetic dependence of the  charged pion mass for the three different cases, including the result using $\chi$PT from \cite{Andersen:2012zc}.
All curves increase with increasing magnetic field, except for the case of constant $m_{ud}$. This result reinforces the 
importance of the magnetic field behavior of $m_{ud}$.

\begin{figure}
\centering
\includegraphics[scale=0.6]{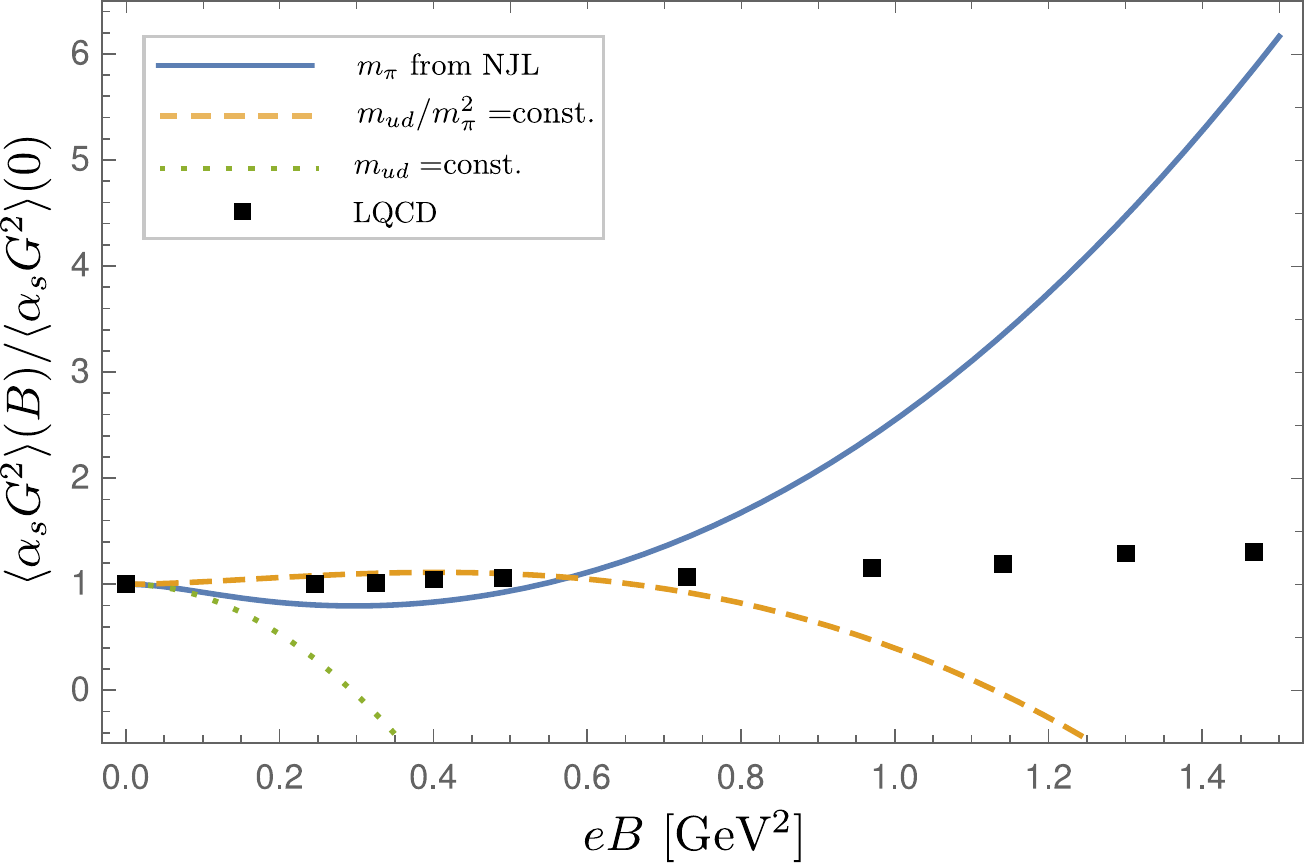}
\caption{\small Magnetic evolution of the gluon condensate considering the three conditions used as an input. 
For comparison, we include the LQCD results from Ref. \cite{DElia:2015eey}}
\label{plot:aGG}
\end{figure}

Finally, we consider the behavior of the gluon condensate. 
This is an interesting parameter because it is not related to  chiral symmetry restoration, but rather to conformal symmetry \cite{DElia:2002hkf, Colangelo:2013ila}.
Figure \ref{plot:aGG} shows the magnetic behavior of the normalized gluon condensate for the three different cases. For a constant quark mass the gluon condensate drops dramatically. This strongly suggests that a constant quark mass is not a valid approximation. In the case where $m_{ud}/m_\pi^2$ is constant, the gluon condensate increases slightly to then decrease gently with increasing magnetic field.
A decreasing  $\langle G^2\rangle$  was found in \cite{Agasian:1999sx}, vanishing at a similar critical value of $eB$. For the pion mass as an input from NJL the gluon condensate starts decreasing  followed by a sharp increase.
The two cases considering magnetic evolution of quark masses show  no important variations for  $eB < 0.7$\,[GeV$^2$].\\

 \section{Conclusions}

In this paper we determined the magnetic behavior of several QCD and hadronic parameters using a set of four FESR. Two sum rules involved the correlator of two axial-vector currents, one involved the axial-vector current together with its  divergence, and another involved two divergences of the axial-vector currents. 
The magnetic field behavior of the chiral condensates was an input from NJL or from LQCD. 
Three different scenarios were considered. The first used the magnetic field dependence of the  pion mass according to NJL results. The second scenario assumed a constant ratio $m_{ud}(eB)/m_\pi^2(eB)$. The third case assumed a constant quark mass, independent of the magnetic field, which can be discarded as concluded from Figs. \ref{plot:mpi} and \ref{plot:aGG}. 
The qualitative magnetic field behavior of $f_\pi$ and $s_0$  appears robust as it results from the first two cases. Regarding the gluon condensate, its behavior above $eB \simeq 0.7$ $\mbox{GeV}^2$ is strongly dependent on whether the pion mass is given by NJL or it is such that $m_{ud}/m_\pi^2 $ is constant. Below this critical magnetic field strength the sum rule results are in good agreement with LQCD. The behavior of the gluon condensate for extreme values of the magnetic field would require further study, beyond the scope of this paper.

\section*{Acknowledgments}

C.V. acknowledge Norberto Scoccola for valuable discussions.
This work was supported by the National Research Foundation (South Africa),
FONDECYT  (Chile) under Grants No. 1170107, No. 1150471, and No. 1150847, and 
Conicyt PIA/BASAL (Chile) Grant No. FB0821.

\end{document}